\documentclass[11pt]{article}

\usepackage{graphics}
\usepackage{amsmath,amssymb,bm,graphicx,url,epsfig}
\usepackage{amsbsy}

\newcommand{\rf}[1]{(\ref{#1})}
\newcommand{\beq}{\begin{equation}}
\newcommand{\eeq}{\end{equation}}
\newcommand{\be}{\begin{equation}}
\newcommand{\ee}{\end{equation}}
\newcommand{\bea}{\begin{eqnarray}}
\newcommand{\eea}{\end{eqnarray}}
\newcommand{\eq}[1]{eq.~(\ref{#1})}
\newcommand{\non}{\nonumber \\*}
\newcommand{\ie}{{i.e.}\ }
\renewcommand{\tilde}{\widetilde}

\newcommand{\e}{\,\mbox{e}}
\renewcommand{\d}{{\rm d}}

\newcommand{\blambda}{\bar\lambda}
\newcommand{\brho}{\bar\rho}
\newcommand{\tr}{\mathrm{tr}}
\newcommand{\LA}{\left\langle}
\newcommand{\RA}{\right\rangle}

\def\fun#1#2{\lower3.6pt\vbox{\baselineskip0pt\lineskip.9pt
\ialign{$\mathsurround=0pt#1\hfil##\hfil$\crcr#2\crcr\sim\crcr}}}

\begin{document}

\hfill  ITEP--TH--27/15
\begin{center}
\vspace{24pt}
{ \large \bf 
String theory as a Lilliputian world
}

\vspace{30pt}

{\sl J. Ambj\o rn}$\,^{a,b}$,   and {\sl Y. Makeenko}$\,^{a,c}$

\vspace{24pt}
{\footnotesize

$^a$~The Niels Bohr Institute, Copenhagen University\\
Blegdamsvej 17, DK-2100 Copenhagen, Denmark.\\
{ email: ambjorn@nbi.dk}\\

\vspace{10pt}

$^b$~IMAPP, Radboud University, \\
Heyendaalseweg 135, 6525 AJ, Nijmegen, The Netherlands\\

\vspace{10pt}

$^c$~Institute of  Theoretical and Experimental Physics\\
B. Cheremushkinskaya 25, 117218 Moscow, Russia.\\
{ email: makeenko@nbi.dk}\\

}
\end{center}
\vspace{48pt}

\begin{center}
{\bf Abstract}
\end{center}

Lattice regularizations of the bosonic string allow no 
tachyons. This has often been viewed as the reason why
these theories have never managed to make any contact
to standard continuum string theories when the dimension
of spacetime is larger than two. We study the continuum
string theory  in large spacetime dimensions
where simple mean field theory is reliable. By keeping 
carefully the cutoff we show that precisely the existence 
of a tachyon makes it possible to take a scaling limit which reproduces
the lattice-string results. We compare this scaling limit with
another scaling limit which reproduces standard continuum-string results.
If the people working with lattice regularizations of 
string theories are akin to Gulliver they will view the standard string-world
as a Lilliputian world no larger than a few lattice spacings.


\newpage

\section{Introduction}

A first quantization of the free particle using the path integral requires a regularization.
A simple such regularization is to use a hypercubic $D$-dimensional lattice if the 
particle propagates in $D$-dimensional spacetime. The allowed wordlines for a 
particle propagating between two lattice points are link-paths connecting the
two points and the action used is the length of the path, i.e. the number of links
of the path multiplied with the link length  $a_\ell$.  
This $a_\ell$ is the UV cutoff of the 
path integral. The lattice regularization works nicely and the limit $a_\ell \to 0$ 
can be taken such that one
obtains the standard continuum propagator.

Similarly, the first quantization of the free bosonic string using the path integral
requires a regularization. It seemed natural to repeat the successful story of the 
free particle and use a  hypercubic $D$-dimensional lattice if the 
wordsheet  of the string lived in $D$-dimensional spacetime, the worldsheet 
being a  connected plaquette lattice surface \cite{lattice}. The  Nambu-Goto action, the area of the worldsheet, would then  be the number of worldsheet 
plaquettes times $a_\ell^2$, $a_\ell$ again denoting the lattice spacing. However, in this case one could not reproduce the results obtained by standard canonical quantization of the string. First, one did not obtain the whole set of 
string masses, starting with the tachyon mass, but only a single, positive mass state.
Next, after having renormalized the bare string coupling constant to obtain a finite lowest 
mass state, this renormalization led to an infinite physical string tension for  strings 
with extended boundaries.

Although it was not clear why the hypercubic lattice regularization  did not work,
the formalism  know as dynamical triangulation (DT) was suggested
as an alternative \cite{DT}. It discretized the independent intrinsic worldsheet geometry
used in the Polyakov formulation of bosonic string theory \cite{Pol81} and the integration
over these geometries were approximated by a summation over triangulations
constructed from equilateral triangles with link lengths $a_t$, where $a_t$ 
again was a UV cutoff. However, the results were identical to the hypercubic
lattice results. In contrast to the hypercubic lattice model the  DT model can be defined when the dimension of spacetime is less than two, where one encounters the so-called
non-critical string theory. This string theory can be solved both using standard continuum 
quantization and using the DT-lattice regularization (and taking the limit $a_t \to 0$).  
Agreement is found. Thus a lattice regularization is not incompatible with string
theory as such. However, in the lattice regularized theories it is impossible to 
have a tachyonic lowest mass state and such states appear precisely when the 
dimension of spacetime exceeds two. It might explain the failure of lattice strings
to connect to continuum bosonic string theory in dimensions $D > 2$. 

The purpose of this article is to highlight the role of the tachyon in connecting 
the continuum bosonic string theory to lattice strings. In \cite{AM15b} we showed how one could technically make such a connection. However, the role 
played in the connection by  the tachyon was not
emphasized to the extent it deserves. 

\begin{figure}[t]
\begin{center}
\includegraphics[width=7cm]{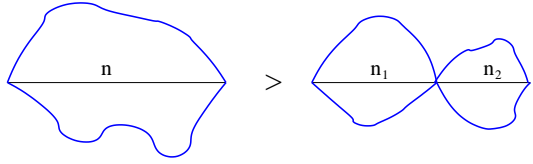}
\end{center}
\caption{Illustration of why $G(n) > G(n_1)G(n_2),~~n=n_1+n_2$.}
\label{fig1}
\end{figure}

Before starting the discussion let us make 
clear why there are no tachyons in, say, a hypercubic lattice string theory 
\cite{lattice,ADJ97}. Consider the two point function for a closed bosonic string on the lattice. We have an entrance loop 
of minimal length, say four links spanning a plaquette (not belonging to the 
string worldsheet) and a similar exit loop, separated by $n$ lattice spacings. The genus zero closed string two-point function $G(n)$ is the sum over all plaquette cylinder lattice surfaces $F$ with these two boundary loops and with $F$ assigned the weight $e^{-\mu N(F)}$, $N(F)$ being the number of plaquettes of  $F$, 
$\mu$ the 
(dimensionless) string tension and $\mu N(F)$  the Nambu-Goto action associated with 
$F$. Clearly this sum is larger than the sum over 
surfaces where the surfaces are constraint to meet at 
a "bottleneck"  (again a single plaquette not belonging to $F$) separated by $n_1$ lattice spacing from the entrance loop, $n_1 < n$ (see Fig. \ref{fig1}), i.e.
\beq\label{jc1}
G(n) \geq  G(n_1) \,G(n-n_1),
\eeq
and thus $-\log G(n)$ is a subadditive function.
Further it can be shown that $G(n) \to 0$ for  $n \to \infty$. 
According to Feteke's lemma this implies that $(-\log G(n))/n$ converges to a real non-negative number for $n \to \infty$ and consequently the lowest mass cannot be tachyonic.

Let us list some other  results obtained in the hypercubic lattice string theory, formulated in dimensionless lattice units.   The theory has a critical point $\mu_c$, such that the partition function is defined for $\mu > \mu_c$ and the 
scaling limit (where one can attempt to define a continuum theory) is obtained for 
$\mu \to \mu_c$. One finds (up to subleading corrections)
\beq\label{janb1}
G_\mu (n) \sim \e^{- m(\mu) n}, \quad m(\mu) \sim (\mu-\mu_c)^{1/4},
\eeq
for $\mu \to \mu_c$. Here $G_\mu (n)$ is the two-point function defined above and  
$m(\mu)$ is the positive mass mentioned above. 
Scaling to a continuum theory is now done by introducing a dimensionful lattice 
spacing $a_\ell$ and by requiring that the two-point function survives when the lattice spacing $a_\ell \to 0$. Thus we write 
\beq\label{janc21}
L = n \cdot a_\ell, \quad m(\mu) \, n = m_{\rm ph} L,\quad \mu \to \mu_c.
\eeq
This determines $a_\ell$ as a function of $\mu$
\beq\label{janc22}
m_{\rm ph}a_\ell(\mu)  = m(\mu) \sim (\mu-\mu_c)^{1/4}.
\eeq
The  problem with the lattice string theory is that the so-called effective string tension
$\sigma(\mu)$ does not scale to zero for $\mu \to \mu_c$ \cite{AD87}. 
The effective string tension is defined as follows (for a closed string):
 compactify one of the lattice directions to $m$ links and insist that the string wraps
 around this dimension once. We still assume that the string propagates
 $n$ links in one of the other lattice directions. Again one can show that the corresponding
 partition function $G_\mu(m,n)$ falls off exponentially with the minimal lattice area 
 $m\times n $ spanned by the string worldsheet:
 \beq\label{jc31}
 G_\mu (m,n) \sim \e^{- \sigma(\mu) \, m\, n}.
 \eeq
 However, $\sigma(\mu)$ does not scale to zero for $\mu \to \mu_c$:
 \beq\label{jc32}
 \sigma(\mu) = \sigma(\mu_c) + c (\mu - \mu_c)^{1/2},~~~~\sigma(\mu_c) > 0.
 \eeq
 Thus the physical, effective string tension $K_{\rm ph}$ (defined analogously 
 to the physical mass $m_{\rm ph}$ in \eq{janc22}) 
 \beq\label{jc33}
 \sigma(\mu) =K_{\rm ph} \, a^2_\ell(\mu)
 \eeq
 scales to infinity for $\mu \to \mu_c$ and $G_\mu(m,n)$ has no continuum limit.     

\section{The bosonic string at large $D$}

Let us consider the closed bosonic string at large $D$. We choose the large $D$
limit because it allows us to perform a reliable mean field calculation, as noticed already
a long time ago \cite{Alv81}. In order to make contact to lattice results we want to control the appearance of the tachyon. We do that by compactifying one of the dimensions
such that it has length $\beta$ and by insisting that the worldsheet wraps once around
this compactified dimension. With this setup there will be no tachyons if only
$\beta$ is larger than the cutoff, as we will show. Also, this setup allows us to define the 
physical, effective string tension precisely as we did it for the lattice string theory.

We use the Nambu-Goto action (to make the situation
analogous to the hypercubic lattice) and deal with the area-action of the embedded 
surface by using a Lagrange multiplier $\lambda^{ab}$
and an independent intrinsic metric $\rho_{ab}$
($\rho :=  \det \rho_{ab} $):
\beq
K_0 \!\int \d^2\omega\,\sqrt{\det \partial_a X \cdot \partial_bX}=
K_0 \!\int \d^2\omega\,\sqrt{ \rho}  
+\frac{K_0}2 \!\int \d^2\omega\, \lambda^{ab} \left( \partial_a X \cdot \partial_bX -\rho_{ab}
\right).
\label{aux}
\eeq
We choose the world-sheet parameters  $\omega_1$ and $\omega_2$ inside 
an $\omega_L\times \omega_\beta$ rectangle in the parameter space and find
the classical solution
\begin{subequations}
\bea
&&X^1_{\rm cl}=\frac L {\omega_L} \omega_1,\quad
 X^2_{\rm cl}=\frac\beta {\omega_\beta} \omega_2,\quad
 X^\perp _{\rm cl} =0,\label{Xcla} \\*
 && 
\left[ \rho_{ab} \right]_{\rm cl}={\rm diag}\left(\frac{L^2}{\omega_L^2},\frac{\beta^2}{\omega_\beta^2}\right),
\label{rhocla} \\*
&& \lambda^{ab}_{\rm cl}={\rm diag}\left(
\frac{\beta\omega_L}{L\omega_\beta},\frac{L\omega_\beta}{\beta\omega_L}\right)
=  \rho^{ab}_{\rm cl} \sqrt{\rho_{\rm cl}},   \label{lacla}
\eea
\label{cla}
\end{subequations}
minimizing the action \rf{aux}.

Quantization is performed using the path integral. We integrate out the quantum fluctuations of the $X$ fields by performing a split
$X^\mu=X^\mu_{\rm cl} +X^\mu_{\rm q}$, where  $X^\mu_{\rm cl} $ is
given by \eq{Xcla}, and then  performing  the Gaussian path
integral over $X^\mu_{\rm q}$. 
We fix the gauge to the so-called static gauge $X_{\rm q}^1=X_{\rm q}^2=0$.%
\footnote{Gauge fixing will in general produce a ghost determinant. 
To leading order in  $D$ this determinant can be
ignored, but it may have to be included in a $1/D$-expansion.}
The number of fluctuating $X$'s is then $d=D-2$. 
We thus obtain the effective action,
governing the fields $\lambda^{ab}$ and $\rho_{ab}$,
\bea
S_{\rm eff}&=& K_0 \int \d^2\omega\,\sqrt{ \rho} 
+\frac{K_0}2 \int \d^2\omega\, \lambda^{ab} \left( \partial_a X_{\rm cl } \cdot \partial_bX_{\rm cl } -\rho_{ab} \right) 
\non &&+ \frac{d}{2}  \tr \log {\cal O}, 
\quad  \quad \quad {\cal O}  := -\frac1 {\sqrt{\rho} }  \partial _a \lambda^{ab} \partial_b.
\label{aux1}
\eea
The operator $-{\cal O}$
 reproduces the usual 2d Laplacian for  $\lambda^{ab}= \rho^{ab}\sqrt{ \rho}$,
and we use the proper-time regularization of the trace
\be
\tr\log {\cal O} 
= - \int_{a^2}^\infty \frac{\d\tau}\tau \tr  \e^{-\tau {\cal O}}, \quad
a^2 \equiv \frac1{4\pi \Lambda^2 }.
\label{5}
\ee

In the mean field approximation which becomes exact at large $D$ 
we disregard fluctuations
of  $\lambda^{ab}$ and $\rho_{ab}$ about the  saddle-point values
$\blambda^{ab}$ and $\brho_{ab} $,
\ie simply substitute them by the mean values, 
which are the minima of the effective action~\rf{aux1}. Integration around the 
saddle-point values will produce corrections which are subleading in $1/D$.

For diagonal and constant 
$\blambda^{ab}$ and $\brho_{ab} $  the explicit formula for the determinant is 
well-known and for $L\gg\beta$ we obtain%
\footnote{The averaged over quantum fluctuations induced metric
 $\LA\partial_a X\cdot \partial _b X \RA$ which equals $\brho_{ab}$ at large $D$
depends in fact on $\omega_1$ near the boundaries, 
but this is not essential for $L\gg \beta$~\cite{AM15b}.}
\bea
S_{\rm eff}&=&\frac{K_0}2 \left( \blambda^{11}\frac{L^2}{\omega_L^2} +
\blambda^{22}\frac{\beta^2}{\omega_\beta^2} +
2\sqrt{\brho_{11}\brho_{22}}  
-\blambda^{11}\brho_{11} -\blambda^{22}\brho_{22}
\right)  \omega_\beta \omega_L \non &&
 -\frac{\pi d}{6}\sqrt{\frac{\blambda^{22}}{\blambda^{11}}}
\frac{\omega_L}{\omega_\beta}-\frac{d\sqrt{\brho_{11}\brho_{22}}\,  \omega_\beta \omega_L}{2\sqrt{\blambda^{11}\blambda^{22}}}
 \Lambda^2.
\label{Sec}
\eea

The minimum of the effective action \rf{Sec} is reached at
 \bea
\brho_{11}&=&\frac{L^2}{\omega_L^2}
\frac{\left(\beta^2-\frac{\beta_0^2}{2C}\right)}{\left(\beta^2-\frac{\beta_0^2}{C}\right)}
\frac{C}{2C-1},\non
\brho_{22}&=&\frac{1}{\omega_\beta^2}\left(\beta^2-\frac{\beta_0^2}{2C}\right)
\frac{C}{2C-1},
\label{rhogen}
\eea
\be
\blambda^{ab} = C\brho^{ab}\sqrt{\brho},\quad C=\frac12 +\sqrt{\frac14-
\frac{d\Lambda^2}{2K_0}},
\label{laws}
\ee
where 
\be
\beta_0^2=\frac{\pi d}{3K_0}.
\ee
Equations \rf{rhogen} and \rf{laws} generalize  the classical solution \rf{cla}.
Note that $C$ as given in \rf{laws} takes values between 
1 and 1/2. This will play a crucial role in what follows. Also note that 
if one were performing a perturbative expansion in $1/K_0$,
 both $C$ and 
eqs.~\rf{rhogen} and \rf{laws} would start out with their classical values.

Substituting the solution~\rf{rhogen} -- \rf{laws} into 
\eq{Sec}, we obtain
\be
S_{\rm eff}^{\rm s.p.} = K_0 C L \sqrt{\beta^2-\beta_0^2/C}
\label{Sfin}
\ee
for the saddle-point value of the effective action, which is nothing but $L$ times
the mass of string ground state.
Further, we find that the average area ${\cal A}=\langle Area \rangle $ of 
the surface which appears in the path integral is 
\be
{\cal A} =
\int \d^2 \omega\,\sqrt{\brho_{11} \brho_{22}} = 
L\frac{\left(\beta^2-{\beta_0^2}/{2C}\right)}{\sqrt{\beta^2-{\beta_0^2}/C}}\frac{C}{(2C-1)}.
\label{72}
\ee

All these results are just a repetition of the original Alvarez 
computation \cite{Alv81}, except that 
he used $\omega_L=L$, $\omega_\beta=\beta$ and, more importantly, he used 
the zeta-function regularization where formally our cutoff $\Lambda$ is put to zero.
As we will see,  maintaining a real cutoff $a$ will be important, so let us 
discuss its relation to a spacetime cutoff  like the hypercubic $a_\ell$ mentioned above.

The cutoff $a$ refers to the operator ${\cal O}$ defined in parameter space and 
provides a cutoff of the eigenvalues of ${\cal O}$. However, the eigenvalues (which
are invariant under change of parametrization) are linked to the target space distances
$L$  and  $\beta$ and {\it not}\/ to the parameters $\omega_L$ and $\omega_\beta$.
Effectively $a$ acts as a cutoff on the wordsheet, measured in length units from 
target space, in agreement with the way we introduced 
$\rho_{ab}$ and $\lambda^{ab}$ in the first place. We can write symbolically
\beq\label{js35}
(\Delta s)^2 = \rho_{ab} \Delta \omega^a \Delta \omega^b = \Delta X \cdot  \Delta X,
\eeq 
where $\Delta s \sim a$ and $\Delta \omega \sim a/\sqrt[4]{\rho}$, which
reflects to what extent the eigenfunctions of 
${\cal O}$ which are not suppressed by the proper-time cutoff $a$ can resolve 
points on the worldsheet. 
This is true semiclassically where the worldsheet is just the minimal 
surface \rf{cla} and it will be true when we consider genuine quantum surfaces which 
are much larger. In the latter case \eq{js35} has to
be averaged over the quantum fluctuations, so
 $\rho_{ab}$ will change accordingly (see \eq{rhogen}), such that the allowed eigenfunctions still can resolve these larger surfaces down
to order $a$, measured in target space length units.  

Formulas \rf{Sfin} and \rf{72} are our main results, valid for $L \gg \beta$ in the 
mean field or large $D$  approximation. The term $-\beta_0^2/C$ appearing 
under the square root in \eq{Sfin} is a manifestation of the closed string 
tachyon as first pointed out in \cite{Arv83,Ole85} 
and this minus-sign will be essential for the limit we take
in the next Section and which will reproduce the lattice string scaling.

\section{The lattice-like scaling limit\label{s:ge}}

Equation  \rf{laws}  shows that the bare
string tension $K_0$ needs to be renormalized in order for $C$ to remain real
since this requirement forces
\beq\label{jc2}
K_0 > 2d \Lambda^2 = \frac{d}{2\pi a^2}.
\eeq
Also, $C$ is clearly constraint to take values between 1 and 1/2
when $K_0$ is decreasing from infinity to $2d\Lambda^2$.
We also require that  $S_{\rm eff}$ is real. This is ensured for all allowed values
of $K_0$ if 
\beq\label{jc3}
\beta^2 > \beta_{\rm min}^2 =\frac{(2\pi a)^2}{3}.
\eeq
For $\beta \geq  \beta_{\rm min}$ we have no tachyonic modes, precisely
the scenario needed if we should have a chance to make contact to lattice string theory.

At first glance it seems impossible to obtain a finite $S_{\rm eff}$ by renormalizing
$K_0$ in \rf{Sfin}, since $K_0$ is of order $\Lambda^2$. However, let us try to 
imitate as closely as possible the calculation of the two-point function 
on the lattice by choosing, for a fixed cutoff $a$ (or $\Lambda$), $\beta$ as small as 
possible without  entering into the tachyonic regime of   $S_{\rm eff}$, i.e.\ by choosing 
$\beta = \beta_{\rm min}$. With this choice we obtain
\beq\label{janx11}
S_{\rm eff}^{\rm s.p.}= 
\sqrt{\frac{\pi}{3}} \; \frac{K_0 CL}{\Lambda} \; \sqrt{2C-1}.
\eeq
Only if $ \sqrt{2C-1} \sim 1/\Lambda$ can we obtain a finite limit for 
$\Lambda \to \infty$.
Thus we are forced to renormalize $K_0$ as follows
\beq\label{janx12}
K_0 = 2d \Lambda^2 +  \frac{\tilde{K}_{\rm ph}^2}{ 2d \Lambda^2}
\eeq
where $\tilde{K}_{\rm ph}$ is finite in the limit $\Lambda \to \infty$.
With this renormalization we find 
\beq\label{janx12a}
S_{\rm eff}^{\rm s.p.} = m_{\rm ph} L,~~~~m_{\rm ph}^2 = \frac{\pi d}{6} \tilde{K}_{\rm ph}.
\eeq
Note that $m_{\rm ph}$ and $\tilde{K}_{\rm ph}$ are proportional to $d$ as to 
be expected in a large $d$ limit.

Since the partition function in this case has the interpretation as a kind
of the two-point function for a string propagating a distance $L$, we have 
the following leading $L$ behavior of the two-point function
\beq\label{janx13}
G(L) \sim \e^{-S_{\rm eff}^{\rm s.p.}}  = \e^{- m_{\rm ph} L} ,
\eeq
where the mass $m_{\rm ph}$ is a tunable parameter.
Note that we have the classical value $C=1$ and
a semiclassical expansion in $1/K_0$ interpolating between $C=1$ and
the quantum value $C=1/2$, which is very similar to the situation for the free particle
where a semiclassical 
expansion in the inverse bare mass interpolates between the classical and quantum cases.

In the scaling limit \rf{janx12} we can calculate the average area 
$ {\cal A}$ of a surface using \rf{72}:
\beq\label{janx14}
 {\cal A}\propto \frac{L}{m_{\rm ph}^3 a^2}.
\eeq
It diverges when the cutoff $a \to 0$. This is to be expected. 
The quantum fluctuations of the 
worldsheet is included in the effective action \rf{Sec} and the same thing will happen 
if we consider a free particle propagating a distance $L$ and 
integrate out the quantum fluctuations. The average length $\ell$  of a quantum path 
in the path integral will be
 \beq\label{jc6}
\ell \propto \frac{L}{m_{\rm ph} a}.
\eeq 
We can express \rf{janx14} and \rf{jc6} in dimensionless units
\beq\label{jc5}
n_L= \frac{L}{a}, \quad \quad n_{\cal A} = \frac{{\cal A}}{a^2} \propto \frac{1}{m_{\rm ph}^3L^3} \; n_L^4, \quad \quad n_\ell = \frac{\ell}{a}  \propto \frac{1}{m_{\rm ph}L} \; n_L^2.
\eeq 
These formulas tell us  the Hausdorff dimension of a quantum surface is $d_H=4$
and the Hausdorff dimension of a quantum path of a particle is $d_H=2$ in the 
scaling limit where $m_{\rm ph}L$  is kept fixed while the cutoff $a \to 0$.
 
 Let us now discuss how we define  the physical string tension.
 With the given boundary conditions the string extends over the minimal
 area $A_{\rm min}= \beta  L$ and we write the partition function as 
 \beq\label{janx20}
 Z(K_0, L, \beta) = \e^{-S_{\rm eff}^{\rm s.p.} (K_0,L,\beta)} = \e^{-K_{\rm ph} A_{\rm min} +
 {\cal O}(L,\beta)}.
 \eeq
 This is precisely the way one would define the physical (renormalized) string tension 
 in a lattice gauge theory via the correlator of two periodic Wilson lines of length
 $\beta$ separated by the distance
 $L\gg\beta \gg a$, where $a$ is the lattice spacing. This is also the way the 
 physical string tension is defined in lattice string theories as discussed in the Introduction.

 From the explicit form of  $S_{\rm eff}^{\rm s.p.}$ given in \rf{Sfin} we have from \rf{janx12}:
 \beq\label{janx22}
 K_{\rm ph} = K_0 C = d \Lambda^2 + \frac{1}{2}\tilde{K}_{\rm ph}+ 
{\cal O}(1/\Lambda^2).
 \eeq
 Thus the physical string tension as defined above diverges as the cutoff
 $\Lambda$ is taken to infinity. However, the first correction is finite and 
 behaves as we would have liked $K_{\rm ph}$ to behave, namely as 
 $\tilde{K}_{\rm ph} \propto m_{\rm ph}^2/d $.  
   
We have thus reproduced the scenario from the lattice strings: it is
possible by a renormalizing of the bare  coupling constant
($K_0=1/(2\pi\alpha'_0)$) to define 
a two-point function with a positive, finite mass. In the limit where the 
cutoff $a \to 0$ the Hausdorff dimension of 
the ensemble of quantum surfaces is $d_H =4$,  but then the effective string tension 
defined as in \eq{janx20} will be infinite. In addition the relation \rf{janx22}
is {\it precisely} the relation \rf{jc32} from the lattice string theories. To make this 
explicit  let us introduce dimensionless variables
\beq\label{janc23}
\mu = K_0 a^2,\quad \quad \mu_c = \frac{d}{2\pi},\quad\quad n = \frac{L}{a},
\quad \quad \sigma(\mu) = K_{\rm ph} a^2.
\eeq
Then the renormalization of $K_0$, \eq{janx12}, can be inverted to define 
the cutoff $a$ in terms of $\mu- \mu_c$ and it becomes identical to \eq{janc22}.
Similarly eqs.~\rf{janx12} and \rf{janx12a} can now be written as
\beq\label{janc24}
m(\mu)\, n = m_{\rm ph} L ,\quad \quad \sigma(\mu) = \sigma(\mu_c) +  
c (\mu-\mu_c)^{1/2},
\eeq
where 
\beq\label{janc25}
m(\mu) \sim (\mu -\mu_c)^{1/4},\quad \quad\sigma(\mu_c) = \frac{\mu_c}{2}  > 0,
\quad\quad c = \frac{1}{2\sqrt{\mu_c}}.
\eeq
Thus one obtains identical scaling formulas by continuum renormalization and by 
lattice renormalization.

\section{Scaling to the standard string theory limit\label{standardS}}

When we integrated out the quantum fluctuations of the worldsheet we made 
decomposition $X^\mu = X^\mu_{\rm cl} + X^\mu_{\rm q}$, 
where the parameters $L$ and $\beta$ refer to the ``background''  fields
$X^\mu_{\rm cl}$. In standard quantum field theory we usually have to 
perform a renormalization of the background field to obtain a finite effective 
action. It is possible to do the same here by rescaling 
\beq\label{jany5}
X_{\rm cl}^\mu = Z^{1/2} X_R^\mu, \quad Z =(2C-1)/C.
\eeq
Notice that the field renormalization $Z$ has a standard perturbative  expansion
\beq\label{jany6}
Z = 1 - \frac{d\Lambda^2}{2K_0} + {\cal O}(  K_0^{-2})
\eeq
in terms of the coupling constant $K_0^{-1} $, which in perturbation  theory 
is always assumed to be small, even compared to the cutoff. 

However, in the limit 
$C \to 1/2$ it has dramatic effects since,
working with renormalized lengths $L_R$ and $\beta_R$ defined as in \rf{jany5}:
\be
L_R=\sqrt{ \frac{C}{2C-1}}\; L,\qquad 
\beta_R=\sqrt{ \frac{C}{2C-1}} \;\beta,
\label{LRbetaR}
\ee
we now obtain for the effective action
\beq\label{jany7}
S_{\rm eff} = K_R \; L_R \sqrt{ \beta_R^2 - \frac{\pi d}{3 K_R}},\quad K_R = 
K_0 (2C-1)\equiv \tilde{K}_{\rm ph}.
\eeq
The renormalized coupling constant $K_R$ indeed makes $S_{\rm eff}$ 
finite and is identical to the $\tilde{K}_{\rm ph}$ defined in \rf{janx12}.
{\it In fact the renormalization $ K_R = (2C-1) K_0$ is identical to the renormalization
\rf{janx12} for $\Lambda \to \infty$.}
 If we view $L_R$ and $\beta_R$ as representing physical distances, \eq{jany7}
 tells us that we have a renormalized, finite  string tension $\tilde{K}_{\rm ph}$ 
 in the scaling limit and even more, \rf{jany7} {\it is}  the  
Alvarez-Arvis continuum string theory formula \cite{Alv81,Arv83}.

The background field renormalization makes the average area ${\cal A}$ of the 
woldsheet  finite. If  the scaling \rf{jany5} for $X_{\rm cl}^\mu$ and \rf{jany7} for
$K_0$ is inserted in the expression \rf{72} for ${\cal A}$ we obtain
\beq\label{jany9}
{\cal A} = L_R \;
\frac{\left(\beta_R^2-\frac{\pi d}{6 K_R}\right)}{\sqrt{\beta_R^2-\frac{\pi d}{3 K_R}}},
\eeq
which  is cutoff independent and thus finite when the cutoff is removed.
The area is simply the minimal area for $\beta_R^2 \gg \pi d/ (3K_R)$  and
diverges when $\beta_R^2 \to \pi d/ (3K_R)$.

\section{Discussion}

As mentioned in the Introduction the fact that lattice string theories seemingly 
are unable to produce anything resembling ordinary bosonic string theory 
has often been ``blamed" on the absence of a tachyonic mass in these regularized 
theories, but it is of course difficult to study the role of the tachyon in a theory
where it is absent.  Here we have addressed the problem from the continuum 
string theory point of view by repeating  the old calculation \cite{Alv81}, while keeping a dimensionful cutoff $a$ explicitly.  In the continuum calculation the tachyonic  term $-\beta_0^2/C$ appears in formula \rf{Sfin}, and  if this term was not negative  
it would be impossible to find a renormalization of $K_0$ which 
reproduces the lattice string scenario.

Somewhat surprising the same renormalization of $K_0$ can produce a completely 
different scaling limit (the conventional string theory limit) provided we are allowed to perform a ``background" renormalization of the coordinates $X_{\rm cl}^\mu$. It is seemingly
difficult to reconcile the two scaling limits. In the limit where the cutoff $a \to 0$ 
we can write \rf{LRbetaR}  as 
\beq\label{jf1}
L = a \cdot \sqrt{ 2 K_R/\mu_c} \, L_R,~~~~
\beta = a \cdot \sqrt{ 2 K_R/\mu_c} \, \beta_R.
\eeq
Thus the scaling limit where $K_R$, $L_R$ and $\beta_R$ are finite as $a \to 0$
is a limit where $L$ and $\beta$ are of the order of the cutoff $a$.
From the point of view of the hypercubic lattice theory we have the lattice cutoff $a_\ell$  which acts simultaneously
as a cutoff in the target space where the string is propagating and as a cutoff on the 
worldsheet  of string. In the lattice world (``Gulliver's world") everything is defined 
in terms of $a_\ell$ and the lattice scaling is such that Gulliver's $L \gg a_\ell$. Since 
we have argued that one essentially  can identify the proper-time cutoff $a$ 
with a minimal
distance $a$ in $R^D$ similar to $a_\ell$, the conventional string limit where $L_R$
and $\beta_R$  are kept fixed becomes a ``Lilliputian world" since $L$ 
and $\beta$ are then of the order  $a_\ell$ from Gulliver's perspective.   
Gulliver's tools are too coarse to deal with the Lilliputian world.

\subsection*{Acknowledgments}

We thank Poul Olesen, Peter Orland and Arkady Tseytlin for valuable 
communication.
The authors acknowledge  support by  the ERC-Advance
grant 291092, ``Exploring the Quantum Universe'' (EQU).
Y.~M.\ thanks the Theoretical Particle Physics and Cosmology group 
at the Niels Bohr Institute for the hospitality. 
The authors also thank Dr. Isabella Wheater for encouragement to use
the Lilliputian analogy.

\end{document}